\documentclass{aa}
\usepackage{epsfig}
\usepackage{graphicx}

\usepackage{epstopdf}
\newcommand{\eqref}[1]{(\ref{#1})}

\begin{document}

\title{Dissipative instability in a partially ionised prominence plasma slab}
\author{  I. Ballai\inst{1,2}, B. Pint\'er\inst{3}, R Oliver\inst{4}  \and M. Alexandrou\inst{1},}
\institute{Solar Physics and Space Plasma Research Centre (SP$^2$RC), Department of
Applied Mathematics, The University of Shef{}field, Shef{}field,
UK, S3 7RH, email: {\tt i.ballai@sheffield.ac.uk}
\and Konkoly Observatory, MTA Research Centre for Astronomy and Earth Sciences, Konkoly-Thege Mikl\'os \'ut 15-17, 1121, Budapest, Hungary
\and Solar System Physics Research Group, Institute of Mathematics, Physics and Computer Sciences, Aberystwyth University, Penglais Campus, UK, email:{\tt b.pinter@aber.ac.uk}
\and Departament de F\'isica, Universitat de les Illes Balears, 07122 Palma de Mallorca, Spain, email:{\tt ramon.oliver@uib.es} 
}
\date{Received dd mmm yyyy / Accepted dd mmm yyyy}

\authorrunning{Ballai et al.}
\titlerunning{Dissipative instability in a partially ionised prominence slab}

\abstract{}
{We aim to investigate the nature of dissipative instability appearing in a prominence planar thread filled with partially ionised plasma in the incompressible limit. The importance of partial ionisation is investigated in terms of the ionisation factor and the wavelength of sausage and kink waves propagating in the slab.}
{In order to highlight the role of partial ionisation, we have constructed models describing various situations we can meet in solar prominence fine structure. Matching the solutions for the transversal component of the velocity and total pressure at the interfaces between the prominence slab and surrounding plasmas, we derived a dispersion relation whose imaginary part describes the evolution of the instability. Results were obtained in the limit of weak dissipation. We have investigated the appearance of instabilities in prominence dark plumes using single and two-fluid approximations . }
{Using simple analytical methods, we show that dissipative instabilities appear for flow speeds that are less than the Kelvin-Helmholtz instability threshold. The onset of instability is determined by the equilibrium flow strength, the ionisation factor of the plasma, the wavelength of waves and the ion-neutral collisional rate. For a given wavelength and for ionisation degrees closer to a neutral gas, the propagating waves become unstable for a narrow band of flow speeds, meaning that neutrals have a stabilising effect. Our results show that the partially ionised plasma describing prominence dark plumes becomes unstable only in a two-fluid (charged particles-neutrals) model, that is for periods that are smaller than the ion-neutral collision time.}
{The present study improves our understanding of the complexity of dynamical processes and stability of solar prominences and the role partial ionisation in destabilising the plasma. We showed the necessity of two-fluid approximation when discussing the nature of instabilities: waves in a single fluid approximation show a great deal of stability. Our results clearly show that the problem of partial ionisation introduces new aspects of plasma stability with consequences on the evolution of partially ionised plasmas and solar prominences, in particular. }

\keywords{Magnetohydrodynamics (MHD)---Sun: filaments, prominences---Sun: magnetic fields---Sun: instability}

\maketitle

\section{Introduction}

The lower part of the solar atmosphere is a perfect example of an environment where temperatures are not high enough for a complete ionisation of the fluid and charged particles and neutrals can coexist due to the collision between them. In particular, solar prominences are regions of cool and dense plasmas where the plasma is not fully ionised and the hydrogen ionisation degree could probably vary in different prominences or even in different regions within the same prominence (Hirayama 1986, Patsourakos and Vial 2002). In this state, the role of neutrals becomes important as the source of momentum transfer between species. Since the neutrals are not controlled by the magnetic field, they flow inside the prominence, preventing the formation of any equilibrium that was not dynamic. Gilbert et al. (2007) found evidence for cross-field diffusion of neutrals that could explain the mass loss in quiescent prominences.

The mathematical description of partially ionised gases is different from the standard magnetohydrodynamic (MHD) approach, as the equations needed to fully describe the state and dynamics of the plasma have to contain corresponding equations for each species. Considerable advancement can be achieved if we suppose that the prominence plasma is made up from hydrogen only. In addition, we assume that the plasma is in ionisation equilibrium, that is the number of ionsation and recombination processes are balanced and the time required for an ion to recombine with an electron (or a neutral atom to ionise through collision) is much shorter than any dynamical time scale under discussion (e.g. periods, damping times, etc.).

Although the partial ionisation of the prominence plasma would mean that each constituent species is treated separately, in the case of ionisation equilibrium and strong correlation of temperature between species, the plasma can still be treated as a single fluid and various quantities in the equations (e.g. density, pressure, etc.) are just the sum of partial components for each species. For periods that are shorter than the ion-neutral collisional frequency, the plasma dynamics has to be described within the framework of two-fluid MHD.

A necessary ingredient in our discussion is the equilibrium flow that has a very strong observational evidence in prominences and nowadays they are observed in various spectral lines, such as H$_{\alpha}$, UV/EUV (Labrosse et al. 2010). Flow speeds range between 5-20 km s$^{-1}$ in quiescent filaments (Lin et al. 2003) to 15-46 km s$^{-1}$ in active region prominences, seen by Hinode/SOT (Okamoto et al. 2007). At the same time, complex dynamics containing vertical upflows and downflows were observed by Berger et al. (2008) in limb prominences. Vorticities inside the prominence of sizes approximately of the order of $10^5 \times 10^5$ km$^2$ were observed by Liggett and Zirin (1984) with rotation rates of 30 km s$^{-1}$.

In our study we assume that the plasmas we are dealing with are non-ideal. According to the standard picture, dynamical processes taking place in these media will be affected by non-ideal mechanisms by lowering or amplifying their amplitude; as a result the energy stored in waves and oscillations are dissipated or accumulated. The plasmas under consideration are dissipative with anisotropic viscosity in the solar corona, while the dynamics in the prominence plasma is affected by dissipative processes that are characteristic for its ionisation state. 

It is well known that the collisions between ions and neutrals introduce an effective anisotropic resistivity into the single fluid equations (Cowling 1957; Braginskii 1965), called Cowling resistivity, which acts only on perpendicular currents. Khodachenko et al. (2004) estimated that in the photosphere and chromosphere the magnitude of this resistivity is a few orders of magnitude larger than the classical parallel Spitzer resistivity, for chromospheric parameters this difference is of the order of $10^4-10^5$. The effect of the Cowling resistivity was investigated in connection to wave damping (Goodman 2001, Leake et al. 2005, Forteza et al. 2008, Carbonell et al. 2010, Singh and Krishan 2010), flux emergence (Leake and Arber 2006, Arber et al. 2007), the formation of nonlinear force-free fields in the chromosphere from photospheric fields (Arber et al. 2009), or even reconnection (Ni et al. 2007).

The problem of the stability of plasmas, and prominences in particular, is a very important aspect of solar atmospheric physics as instabilities can disrupt the magnetic configuration on large scales or can generate turbulent convection cells, which significantly enhance the transport of energy across magnetic surfaces. In general, these instabilities can be categorised in many ways. In the present study, however, we can classify them as ideal instabilities and non-ideal instabilities. In the first class, the source of instability (the reservoir for energy and momentum gain and growth) comes from currents or pressure gradients that are present in ideal plasmas. Particular examples of ideal instabilities are the Kelvin-Helmholtz (KH) and Rayleigh-Taylor (RT) instabilities, that have been studied in detail in hydrodynamics and MHD (e.g.  Soler et al. 2012,  Murawski et al. 2016, Oliver et al. 2016, etc.  studied the triggering and evolution of the KH instabilities in prominence plasmas, while Diaz et al. 2012, Shadmehri et al. 2013, Diaz et al. 2014, Khomenko et al. 2014b, Ruderman 2015, etc. discussed the problem of RT instability in solar prominences).

On the other hand, non-ideal instabilities assume that the plasma contains some sort of transport mechanism that helps in the triggering of instability (e.g. dissipative instability, thermal instability, tearing mode instability, etc.). In these instabilities the presence of a non-ideal effect is a necessity. For instance, Hillier et al. (2010) studied the effect of Cowling resistivity on the Kippenhahn and Schluter prominence model. They found that due to the inclusion of the Cowling resistivity, the tearing mode instability time scale is reduced by more than one order of magnitude, meaning that the structure of the whole prominence can be significantly altered by the Cowling resistivity. Recently Mart\'inez-G\'omez et al. (2015) studied the onset of the KHI in a two fluid plasma where non-ideal effects due to collision between species was considered. They found that due to the collision between ions and neutrals the KHI can appear even for sub-
Alfv\'enic flow speeds and these collisions are able to reduce the growth rate of unstable perturbations but they cannot stop the instability completely. In an earlier paper, Ballai et al. (2015) studied the appearance of dissipative instability at the interface of prominence and corona plasmas assuming that the prominence is partially ionised (they assumed the Cowling resistivity to be important) and the coronal plasma is viscous. 

The present study is a normal extension of their previous work. Here we assume that the magnetic field permeating the plasma is structured, an assumption that brings the description of the instability phenomenon closer to a real situation. The onset of dissipative instability is discussed for two configurations. In order to emphasise the importance of a two-fluid MHD, we discuss the same problem in two different descriptions corresponding to two different regimes compared to the ion-neutral collisional time. The paper is structured as follows. In Section 2, we introduce the equilibrium configuration and the mathematical formalism that will be used to determine the dispersion relation of waves and ultimately the threshold at which instabilities occur. Section 3 is dedicated to the derivation of the dispersion relations of waves propagating in the magnetic structure. This is obtained by matching the solutions at the interfaces. In section 4, we discuss the generation of the dissipative instability for all the models used in the study, differentiating between the results obtained in a single and two-fluid approximation. Finally, we conclude and summarise our results in Section 5.

\section{The equilibrium configuration}

The magnetic field in the solar atmosphere tends to accumulate into entities (flux tubes, coronal loops) of finite size (radius) and very often this size is determined by the balance of various forces acting upon these structures. Once waves will propagate in finite size waveguides, their phase speed becomes dependent on the wavelength at which they propagate, that is they become dispersive (in optics this phenomenon is also called waveguide dispersion). Depending on the particular dependence of the phase velocity on the wavelength, we can differentiate between positive and negative dispersion. In the first case, waves will propagate faster with increasing wavelength, while in the case of waves with negative dispersion, an increasing wavelength would mean a decreasing phase speed. 

In the present study, the structuring of the magnetic field is modelled by a three-layer model, where a magnetic slab along the $x$-axis of thickness $z_0$ is sandwiched between two semi-infinite planes situated at $z=0$ and $z=z_0$. The magnetic field in the three regions is taken to be parallel to the $x$ axis, that is ${\bf B}_0=B_0{\hat {\bf x}}$, where ${\hat {\bf x}}$ is the unit vector in the $x$ direction. Depending on the possible roles played by different regions in the equilibrium configuration, we can study the appearance of dissipative instability in two different models. In the first model, the equilibrium setup describes the case of a partially ionised prominence slab fibril in steady state, which is immersed into the fully ionised and viscous solar corona. The second model consists of a partially ionised prominence fibril sandwiched by an interfibril partially ionised prominence plasma. In both models we assume that inside the slab the plasma flows with a piecewise constant flow in the direction of the magnetic field. In all models studied here, the quantities describing the state of the plasma in the external medium (i.e. outside the slab) are labelled by an index '2', while, inside the slab, the plasma is described through quantities with an index '1'. For simplicity, we assume that the fluids in the two regions (and all models) are incompressible, a generalisation of this restriction would be a rather straightforward task. 

A discussion on the nature of transport processes that can act under these conditions is presented in an earlier paper by Ballai et al. (2015). Here we are going to employ the same considerations and assume that the viscosity is the dominant transport mechanism in the solar corona and it is described by the first term in Braginskii's viscosity tensor, while the transport mechanisms in the prominence appear in the induction equation (single fluid description), or due to the friction of different species of particles evidenced in the momentum equation (two fluid description). The two frameworks describe the same physics, however, applicable for different regimes, depending on how the wave periods compare to the ion-neutral collisional time. 

The relative densities of neutrals and ions are defined as (Forteza et al. 2007)
\begin{equation}
\xi_i=\frac{\rho_i}{\rho}\approx \frac{n_i}{n_i+n_n}, \quad \xi_n=\frac{\rho_n}{\rho}\approx \frac{n_n}{n_i+n_n},
\label{eq:3.8}
\end{equation}
where $\rho_i$ and $\rho_n$ are the mass density of ions and neutrals, $\rho$ is the total density, and $n_i$ and $n_n$ are the number density of ions and neutrals, respectively. The degree of ionisation can be characterised by the ionisation fraction defined as
\begin{equation}
\mu=\frac{1}{1+\xi_i}.
\label{eq:3.9}
\end{equation}
According to this definition, a fully ionised gas corresponds to $\mu=0.5$, while a neutral gas is described by $\mu=1$. The last two relations allow us to express the relative densities in terms of the ionisation degree as
\begin{equation}
\xi_i=\frac{1}{ \mu}-1, \quad \xi_n=2-\frac{1}{ \mu}.
\label{eq:3.10}
\end{equation}
The equations of ideal fully ionised plasma can be obtained by taking $\xi_n=0$ and $\xi_i=1$.

Let us assume that the two-dimensional dynamics in the partially ionised plasma slab is described first within the framework of a single-fluid MHD. Here perturbations in velocity and magnetic field are ${\bf v}=(v_x,0,v_z)$ and ${\bf b}=(b_x,0,b_z)$.  The dynamics in the prominence (inside the slab) is described by the system of non-ideal linearised MHD equations
\begin{equation}
\nabla \cdot {\bf v_1}=0,\quad \nabla \cdot {\bf b_1}=0,
\label{eq:1.1}
\end{equation}
\begin{equation}
\rho_{01}\frac{\partial {\bf v_1}}{\partial t}+\rho_{01}v_0\frac{\partial {\bf v_1}}{\partial x}=-\nabla P_1+\frac{B_{0}}{\mu_0}\frac{\partial {\bf b_1}}{\partial x},
\label{eq:1.2}
\end{equation}
\begin{equation}
\frac{\partial {\bf b_1}}{\partial t}+v_0\frac{\partial {\bf b_1}}{\partial x}=B_{0}\frac{\partial {\bf v_1}}{\partial x}+{{\bf \cal R}},
\label{eq:1.4}
\end{equation}
where $\mu_0$ is the permeability of free space and ${{\bf \cal R}}$ is the resistive term, given by
\[
{{\bf\cal R}}=\eta\nabla^2{\bf b}_1-\Xi\nabla\times (\nabla p_1\times {\bf B}_0)+
\]
\begin{equation}
+\frac{(\eta_C-\eta)}{|{\bf B}_0|^2}\nabla\times\left\{\left[\left(\nabla\times {\bf b_2}\right)\times {\bf B}_0\right]\times {\bf B}_0\right\}.
\label{eq:1.5}
\end{equation}
In the above equation $\eta$ and $\eta_C$ denote the standard Spitzer and Cowling resistivities, respectively and the connection between them is given by
\[
\eta_C=\eta+\frac{\xi_n^2B_0^2}{\mu_0\alpha_n}=\frac{1}{\mu_0}\left[\frac{m_e}{n_e e^2}\left(\frac{1}{\nu_{ie}}+\frac{1}{\nu_{en}}\right)+\frac{\xi_n^2B_0^2}{\alpha_n}\right],
\]
where $\nu_{ei}$ and $\nu_{in}$ are the electron-ion and electron-neutral collisional frequencies and $\alpha_n$ is the friction coefficient given by
\[
\alpha_n=2\xi_n(1-\xi_n)\frac{\rho_0^2}{m_n}\sqrt{\frac{k_BT_0}{\pi m_i}}\Sigma_{in},
\]
and $\Sigma_{in}\approx 5\times 10^{-19}$ m$^{2}$ is the ion-neutral collisional cross section. In the generalised induction equation (\ref{eq:1.4}) we neglected other effects, such as the Biermann's battery term (under solar atmospheric conditions this is too small) and the Hall effect. In addition, in Eq. (\ref{eq:1.2}), $P_1=p_1+B_0b_{x1}/\mu_0$ is the total (kinetic and magnetic) pressure, and the quantity $\Xi$ in Eq. (\ref{eq:1.5}) is given by
\[
\Xi=\frac{\xi_n^2\xi_i}{(1+\xi_i)\alpha_n}.
\] 
It should be noted that the second term in Eq. (\ref{eq:1.5}), called diamagnetic current term, in the limit of weak dissipation will play no role in our further analysis.

Let us assume harmonic oscillations, meaning that perturbations are chosen to be proportional to $\exp[i(kx-\omega t)]$. It was shown earlier by Ballai et al. (2015) that, inside the slab, the total pressure, $P_1$, and $z$ component of the velocity vector, $v_{z}$, are connected by
\begin{equation}
(\Omega+i\eta_{C1}k^2)P_1=\frac{i\rho_{01}}{k^2}(D_{A1}+i\eta_{C1}\Omega k^2)\frac{dv_z}{dz},
\label{eq:1.6}
\end{equation}
where $\Omega=\omega-kv_0$ is the Doppler-shifted frequency, $D_{A1}=\Omega^2-k^2v_{A1}^2$ and $\eta_{C1}$ is the coefficient of the Cowling resistivity in region 1.
In the above calculations we assumed, for simplicity, that in the dissipative terms we can consider $d^2/dz^2\ll k^2$. This simplification is fully justified as the plasma movement takes place in the transversal direction following the oscillatory motion of the Alfv\'enic wave (the plasma is incompressible).

In model 1 (prominence slab surrounded by fully ionised viscous corona), the governing equation for perturbations in the viscous and fully ionised solar corona (see Ballai et al. 2015) is
\begin{equation}
P_2=\frac{i\rho_{02}(D_{A2}+2i\nu k^2\omega)}{k^2\omega}\frac{dv_z}{dz},
\label{eq:1.7}
\end{equation}
where $\nu=\eta_0/\rho_{02}$ is the kinematic coefficient of viscosity and $\eta_0$ is the coefficient of the first term Braginskii's viscosity tensor (for details see, e.g. Ballai et al. 2015).

For the second model (partially ionised plasma slab surrounded by another partially ionised interfibril prominence plasma in a different state of ionisation), the governing equation is 
\begin{equation}
(\omega+i\eta_{C2}k^2)P_2=\frac{i\rho_{02}}{k^2}(D_{A2}+i\eta_{C2}\omega k^2)\frac{dv_z}{dz},
\label{eq:1.8}
\end{equation}
where $D_{A2}=\omega^2-k^2v_{A2}^2$ and $\eta_{C2}$ is the Cowling resistivity in region 2. 

The solutions of these equations must be connected at the boundaries of the regions. We will be concerned with those disturbances that are laterally evanescent, that is $v_z(z)\to 0$ as $|z|\to \infty$, meaning that the energy of the disturbance is essentially confined to the interior of the slab. As a result, the $z$-component of the velocity can be given as (for details see, e.g. Edwin and Roberts 1982, Ballai et al. 2015)
\begin{equation}
v_{z}=\left\{\begin{array} {ll} \beta_ee^{-k(z-z_0)} & \mbox{for}\quad z>z_0,\\ \alpha_0\cosh(kz)+\beta_0\sinh(kz)            
& \mbox{for} \quad 0<z<z_0,\\\alpha_e e^{kz} & \mbox{for}\quad z<0.
\end{array}\right.
\label{eq:1.10}
\end{equation}
The coefficients $\alpha_0$, $\beta_0$, $\alpha_e$, $\beta_e$ are real constants that can be determined after joining the solutions at the boundaries of the slab. According to the standard classification, the only modes that can appear in this structure are surface modes that could be sausage or kink, depending whether $v_z$ is an odd or even function of $z$. 

Given the particular orientation of the equilibrium magnetic field the interfaces between the prominence slab and its environment can be considered as tangential discontinuities. Let us assume that the equation of the perturbed discontinuity is $\zeta(x,t)=0$. The requirements that the normal component of the velocity and normal component of the stress tensor are continuous imply that the linearised kinematic boundary condition reduces to
\[
\left[\left[v_z-{\bf v}_0\cdot \nabla\zeta\right]\right]=0,
\label{eq:1.11}
\]
while the continuity of the stress tensor in the case of homogeneous background is $\left[\left[P\right]\right]=0$ and $[[g]]$ denotes the jump of quantity $g$ across the discontinuity in the sense that the jump of a function $g(z)$ is defined as $[[g(z)]]=\lim_{\epsilon\to 0_{+}}\left[g(z+\epsilon)-g(z-\epsilon)\right]$. The $z$-component of the velocity and $\zeta$ can be related by 
\[
v_z=\frac{\partial \zeta}{\partial t}+{\bf v}_0\cdot\nabla \zeta.
\]
Using the above property, the kinematic boundary condition becomes
\begin{equation}
\left[\left[\frac{v_z}{\omega-{\bf k}\cdot {\bf v}_0}\right]\right]=0.
\label{eq:1.14}
\end{equation}
In the case of model 2 (single fluid approximation), the continuity of the stress tensor simplifies to the requirement that the total pressures on the two sides of the discontinuity are equal, however, in the case of model 1 the viscosity of the corona is modifying this requirement, so that the continuity of the normal component of the stress tensor reduces to
\begin{equation}
P_1=P_2-2\rho_{02}\nu\frac{\partial v_{z2}}{\partial z},
\label{eq:1.15}
\end{equation}
that has to be evaluated at the interfaces, situated at $z=0$ and $z=z_0$. 

\section{Dispersion relation of surface waves propagating in the slab}

Let us first deal with one fluid approximation. In our derivation we assumed that all perturbations oscillate with the same frequency $\omega$, which is a complex quantity that can be written as $\omega=\omega_r+i\omega_i$. 
Let us introduce the viscous and resistive Reynolds numbers as
\begin{equation}
 R_{r}=\frac{v_{A1}}{k\eta_C}, \quad R_v=\frac{v_{A2}}{k\nu}.
\label{eq:2.1}
\end{equation}
Under coronal and prominence conditions, both Reynolds numbers are very large and therefore waves will propagate with little damping over a period, meaning that, in our subsequent calculations, we will consider that $|\omega_r|\gg|\omega_i|$. The very large Reynolds numbers also allow us to consider dissipative terms much smaller than other terms belonging to ideal MHD, meaning that in our calculations all terms containing $\nu^2$ or $\eta_C^2$ are neglected. The interaction of flows and waves in a dissipative medium will generate the new physics our study deals with. Later we will see that, contrary to our expectations, dissipation does not always act towards decreasing the wave amplitude; for specific values of flow speed or ionisation degree, the interplay between flows, dissipation, and waves could lead to an increase of the waves' amplitude, or an unstable behaviour.

The dispersion relation of waves propagating in the magnetic slab can be obtained by imposing the boundary conditions on the total pressure and normal component of velocity. For the $\sinh$ term (see Eq. \ref{eq:1.10}), the dispersion relation of sausage waves reads
\begin{equation}
d\left(D_{A1}+\frac{i\eta_Ck^4v_{A1}^2}{\Omega}\right)+(D_{A2}+4i\nu k^2\omega)\tanh(kz_0)=0,
\label{eq:2.2}
\end{equation}
while the $\cosh$ term leads to the dispersion relation of kink waves 
\begin{equation}
d\left(D_{A1}+\frac{i\eta_Ck^4v_{A1}^2}{\Omega}\right)+(D_{A2}+4i\nu k^2\omega)\coth(k z_0)=0.
\label{eq:2.3}
\end{equation} 
We can rearrange these relations in the form
\[
D_{A2}\left\{\begin{array}{l}\tanh(kz_0)\\ \coth(kz_0)\end{array}\right\}+dD_{A1}+
\]
\begin{equation}
+ik^2\left[4\nu\omega\left\{\begin{array}{l}\tanh(kz_0)\\ \coth(kz_0)\end{array}\right\}+\frac{\eta_Ck^2v_{A1}^2d}{\Omega}\right]=0.
\label{eq:2.4}
\end{equation}
Following the same consideration and imposing the right boundary conditions at the two interfaces, the dispersion relation for the second model becomes
\[
D_{A2}\left\{\begin{array}{l}\tanh(kz_0)\\ \coth(kz_0)\end{array}\right\}+dD_{A1}+ik^4\times
\]
\begin{equation}
\times \left[\frac{\eta_{C1}v_{A1}^2d}{\Omega}+\frac{\eta_{C2}v_{A2}^2}{\omega}\left\{\begin{array}{l}\tanh(kz_0)\\ \coth(kz_0)\end{array}\right\}\right]=0,
\label{eq:2.5}
\end{equation}
where $d=\rho_{01}/\rho_{02}$ is the density contrast between the interior and exterior in the slab. These dispersion relations will be investigated analytically and numerically in the next section to determine the range of flows and thickness of the slab (or wavelength of waves compared to the geometrical transversal size of the slab) for which the incompressible surface waves propagating in the slab are unstable.

\section{Dissipative instability}

Since the Reynolds numbers, defined by Eqs. (\ref{eq:2.1}), are very large, it is realistic to assume that the damping of waves propagating in the magnetic slab is weak. According to the chosen dependence of variables with time, a perturbation with $\omega_i>0$ corresponds to an instability, that is when the amplitude of waves grows exponentially with time according to $\exp(\omega_i t)$. Here we restrict ourselves to the linear phase of instabilities. Linear growth rates provide us with characteristic time scales for the instability to operate. Nonlinear studies are needed to assess the real impact of the instability on the evolution of the plasma parameters. This topic, however, would require numerical analysis, which would be outside the scope of the present study.

Following the method developed by Cairns (1979), the imaginary part of the frequency can be calculated as
\begin{equation}
\omega_i\approx -\frac{{\cal D}_I}{\partial {\cal D}_R/\partial \omega},
\label{eq:3.1}
\end{equation}
where ${\cal D}_R$ and ${\cal D}_I$ are the real and imaginary parts of the dispersion relations (see Eqs. \ref{eq:2.4}--\ref{eq:2.5}) and this expression should be evaluated at the value of the frequency that corresponds to the solution of the real part of the dispersion relation, that is a root of the equation ${\cal D}_R=0$. 

Let us first concentrate on the sausage modes, the solution for kink modes being easily generated. The difference in the two models resides only in the choice of the transport mechanism, therefore it is obvious that the real part (corresponding to the ideal case) will be identical. In this case it is straightforward to show that the root of the real part of the dispersion relation becomes
\[
{\omega}_{\pm}=\frac{k}{d+\tanh kz_0}\left[dv_0\pm \sqrt{d\tanh kz_0(v_{KH}^2-v_0^2)}\right]=
\]
\begin{equation}
\frac{dkv_0}{d+\tanh kz_0}\pm\frac{k\Gamma}{d+\tanh kz_0},
\label{eq:3.2}
\end{equation}
where we used $\Gamma=\sqrt{d\tanh kz_0(v_{KH}^2-v_0^2)}$, and $v_{KH}$ is the Kelvin-Helmholtz speed for propagation in the slab, defined here as
\begin{equation}
v_{KH}^2=\frac{(d+\tanh kz_0)(v_{A1}^2d+v_{A2}^2\tanh kz_0)}{d\tanh kz_0}.
\label{eq:3.3}
\end{equation}
This speed plays a special role in the determination of the nature of instabilities that can appear in the magnetic slab. First of all, Eq. (\ref{eq:3.2}) shows that Kelvin-Helmholtz instabilities (KHI) appear only for those flows that are greater than $v_{KH}$. When waves are restricted to propagate in the slab, even $v_{KH}$ is dispersive and it varies not only with the density ratio (as in the case of wave propagation along a density interface) but also with the relative magnitude of the wavelength compared to the transversal size of the slab. 

Given the importance of $v_{KH}$, it is instructive to estimate the magnitude and variation of this quantity for the two models. Observations show that the wavenumber of waves in prominences varies between $10^{-8}$ and $10^{-6}$ m$^{-1}$ (Forteza et al. 2007). As a characteristic Alfv\'en speed in the prominence, we choose $v_{A1}=28$ km s$^{-1}$ (see Joarder and Roberts 1992). For model 1, we assume three values of the external Alfv\'en speed (198 km s$^{-1}$, 280 km s$^{-1}$, 343 km s$^{-1}$) that - under the assumption of identical field strength - would result in a density contrast of 50, 100 and 150, respectively.  For model 2, we assume that our setup describes the situation of a dark plume, where the internal Alfv\'en speed is $v_{A1}=200$ km s$^{-1}$, that is surrounded by the prominence with a density contrast of $d=0.05$, $0.1$ and $0.5$, respectively, resulting in Alfv\'en speeds of 44, 63 and 141 km s$^{-1}$, respectively. 

A key parameter in our discussion is the product $kz_0$, where $k$ is the wavenumber of the waves under study and $z_0$ is the width of the slab. Since our analysis refers to two possible scenarios (prominence slab surrounded by coronal plasma and prominence slab surrounded by prominence plasma), the value of this parameter ($kz_0$) takes different values. In the first case we are going to assume (hypothetically) that the entire prominence can be considered as one single plasma slab, in which case we are going to consider that $z_0$ is the width of the prominence. The typical width of prominences varies between 4 and 30 Mm (Lin 2010), meaning that the product $kz_0$ falls in the interval $0.01$ and $30$. For the second scenario, we are going to consider that $z_0$ refers to the size of a thread only, that has a width of     
100-600 km, meaning that the dimensionless parameter $kz_0$ will be in the interval $0.001$ and $0.6$. 

One key aspect to note is that regardless of the model employed, the KH speed is always super-Alfv\'enic. Under prominence conditions, these speeds amounts to values that are of the order of a few hundred km s$^{-1}$. This would also mean that, in prominences, the plasma is always Kelvin-Helmholtz  stable.
\begin{figure}
\centering
\includegraphics[width=\columnwidth]{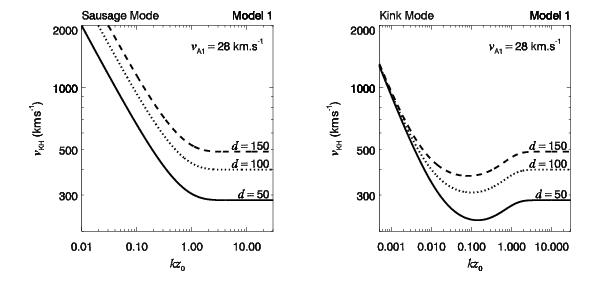}
\caption{Variation of the Kelvin-Helmholtz speed, $v_{KH}$, with the dimensionless quantity $kz_0$ for model 1 on logarithmic scale for three different values of the density contrast between the solar prominence and surrounding solar corona.}
\label{fig1}
\end{figure}
The variation of the threshold speed at which waves propagating in the slab become KH unstable is shown, on logarithmic scale, in Fig. 1 for model 1 with the threshold increasing with the wavelengths for both sausage and kink modes. For both types of waves the range of speed obtained clearly show that the existence of flows larger than $v_{KH}$ are not possible to observe, meaning that the prominence in this model is indeed KH stable. For large values of $kz_0$ (wide slab or long wavelength approximation) the value of the Kelvin-Helmholtz speeds reaches the value obtained for a single interface (see, e.g. Ballai et al. 2015). It is also clear that the threshold were waves become KH unstable increases with the density contrast between the prominence and the solar corona.
\begin{figure}
\centering
\includegraphics[width=\columnwidth]{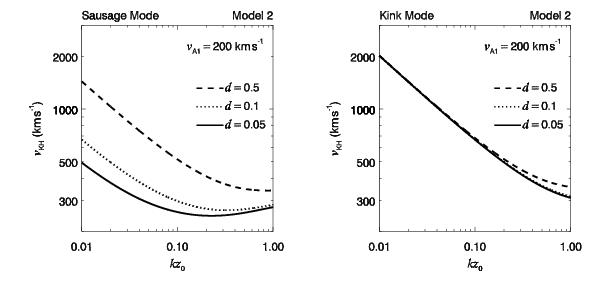}
\caption{Same as Fig. 1, but for model 2}
\label{fig3}
\end{figure}

In the case of model 2, the range of $kz_0$ is different and observations restrict us to the situation when the wavelength of waves is larger than the width of the slab (see Fig. 2). In the case of sausage waves, $v_{KH}$ shows a minimum in the $kz_0\ll 1$ domain (thin slab) that is attained for $kz_0=\tanh^{-1}v_{A1}d/v_{A2}$ and here the value of the $v_{KH}=v_{A1}+v_{A2}$. For kink waves the KH threshold shows a $1/kz_0$-type monotonic decrease. For small values of the dimensionless quantity $kz_0$, the KH threshold for kink waves is much larger than the corresponding value for sausage modes, while they tend to become equal for $kz_0\approx 1$. This result shows that, for long wavelengths, sausage waves can become much easily KH unstable than kink waves, however, the range of speeds obtained here is inconsistent with the values observed for background flows.

Now using the definition of $\omega_i$ together with the dispersion relations (\ref{eq:2.4})--(\ref{eq:2.5}) we obtain that the imaginary part of the frequency in the first model is given by
 \[
 \omega_{i1}=\mp\frac{k^2}{2\Gamma}\left[\frac{4\nu\tanh kz_0(dv_0\pm \Gamma)}{d+\tanh kz_0}-\right.
 \]
 \begin{equation}
\left.- \frac{\eta_Cv_{A1}^2d(d+\tanh kz_0)}{v_0\tanh kz_0 \mp\Gamma}\right],
 \label{eq:3.5}
 \end{equation}
 where the Cowling resistivity in the solar prominence is given by
 \begin{equation}
 \eta_C=10^9T_p^{-3/2}+\frac{(2\mu-1)v_{A1}^2m_p}{2(1-\mu)\rho_1\Sigma_{in}}\left(\frac{\pi m_p}{k_BT_p}\right)^{1/2},
 \label{eq:3.6}
 \end{equation}
where $T_p$ is the temperature in the prominence. Here we assumed that the plasma is made up of hydrogen and, therefore, the mass of ions is equal to the mass of protons. We need to note here that a positive imaginary part of the frequency would mean that the amplitude of waves will grow despite the presence of dissipation, that is these waves are undergoing a dissipative instability.

For model 2, following the same technique, the imaginary part of the frequency becomes
\[
\omega_{i2}\approx \mp \frac{k^2(d+\tanh kz_0)}{2 \Gamma}\left(-\frac{\eta_{C1}v_{A1}^2d}{v_0\tanh kz_0\mp \Gamma}+\right.
\]
\begin{equation}
\left.+ \frac{\eta_{C2}v_{A2}^2\tanh kz_0}{dv_0\pm \Gamma}\right).
\label{eq:3.7}
\end{equation}

Now let us investigate graphically the regions where the plasma becomes unstable, that is we search for the combination of physical parameters that make the imaginary part of the frequency positive. As we specified earlier, the flows that are currently observed in solar prominences are of the order of a few tens of km s$^{-1}$. In the case of model 1 we first plot the frequency contour plot of backward propagating waves (see Fig. 3) showing the regions where the imaginary part of the frequency is changing sign for a given value of the ionisation factor, $\mu=0.95$ (here and thereafter we are going to concentrate mainly on backward propagating waves as forward propagating waves will have a standard physical damping).
\begin{figure}
\centering
\includegraphics[width=\columnwidth]{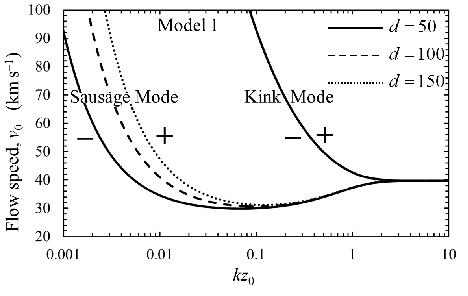}
\caption{Contour plot of the variation of $\omega_{i1}$ in the case of sausage (left-hand side curves) and kink (right-hand side curves) modes in terms of background equilibrium flow and the value of the dimensionless parameter $kz_0$ for model 1. The ionisation rate is $\mu=0.95$.}
\label{fig4}
\end{figure}
The  region above each curve corresponds to a combination of parameters that makes the imaginary part of the frequency positive, meaning that backward propagating waves are unstable. For values of equilibrium flows that are closer to observed values (the lower end of the flow interval considered here), it is possible to obtain two values of $kz_0$ where $\omega_{i1}$ is changing sign, that is the domain of $kz_0$ where waves are unstable is bounded by the two values. For example, for $d=50$ and $v_0=35$km s$^{-1}$,  $\omega_{i1}>0$ for $0.009338<kz_0<0.6246$. The plots in Fig. 3 were obtained for three different values of densities contrast and it is obvious that the threshold value for sausage modes depends on the value of density contrast only for larger values of flows. In the case of kink modes the three curves are indistinguishable, so the instability threshold does not show any dependence on the density contrast. This behaviour could be explained in terms of the internal motion of the plasma in the two wave modes. In the case of kink waves, the slab oscillates without disturbing the internal structure of the slab, while, in the case of sausage modes, the internal plasma structure is compressed and relaxed according to the oscillating pattern of the wave. While the instability of sausage modes sets in for smaller values of $kz_0$ (in the long wavelength limit), kink waves become unstable only when their wavelength is comparable with or shorter than the width of the slab.
  
\begin{figure}
\centering
\includegraphics[width=\columnwidth]{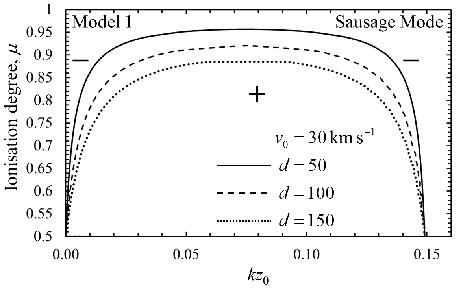}
\caption{Contour plot of the variation of $\omega_{i1}$ in the case of sausage modes in terms of ionisation degree and the value of the dimensionless parameter $kz_0$ for model 1. Here $v_0=30$ km s$^{-1}$. The region where instability occurs is shown by the {\it plus} symbol.}
\label{fig5}
\end{figure}
Let us investigate how the instability threshold varies with the ionisation degree of the prominence slab. We choose a particular value of the equilibrium flow of 30 km s$^{-1}$ and let $\mu$ vary between 0.5 and 1, corresponding to the ionisation state of the plasma (see Fig. 4). We also fix three values of density contrast (d=50, 100, 150)  between the prominence and solar corona. In the case of sausage modes, the threshold of instability depends on the ionisation degree for very limited interval of $kz_0$. For this particular value of flow and density contrast, the backward propagating wave is unstable only for wavelengths that are larger than the width of the slab, in particular $kz_0<0.15$. In addition significant dependence on the ionisation degree occurs only near the end of the interval. Density differences between the two media also influence the instability threshold. Fig. 4 shows that, as the density contrast increases, the threshold moves towards the direction of increased ionisation. The corresponding plot for kink waves is not shown here, as the appearance of unstable modes involves very high values of the equilibrium flow, therefore the appearance of instability is unrealistic.

Let us now discuss the second model that we use to study the instability of waves propagating in prominence dark plumes. Observations by Berger et al. (2008, 2011) and Ryutova et al. (2010) revealed that dark plumes are turbulent upflows in prominences which usually develop Kelvin-Helmholtz vortex rolls. Ca II absorption lines in prominence plumes show these as dark features, in contrast to the prominence material, which suggests a hotter plasma in the plumes compared to their environment. Plumes are also less dense than their surrounding material. The width of plumes ranges between 0.5 Mm to 6 Mm and their maximum heights are between 11 Mm and 17 Mm. The mean flow speed is about 15 km s$^{-1}$, although velocities up to 30 km s$^{-1}$ are also measured, while the typical plume lifetime is between 400 s and 890 s (Berger et al. 2010). 

In this model the plasma inside and outside the slab are partially ionised and the plasma inside the slab exhibits an equilibrium flow along the background magnetic field. These structures are hotter and less dense than their environment, therefore $d<1$. We assume that the plasma in the plume is nearly completely ionised, therefore we choose $\mu_1=0.55$. The ionisation degree of the prominence (region 2) is unknown. We let $\mu_2$ vary in the interval 0.55-0.95. Let us first discuss the sausage modes appearing in these structures. The parameter domains where the imaginary part of the frequency of backward propagating sausage modes changes sign are shown in Fig. 5.
\begin{figure}
\centering
\includegraphics[width=\columnwidth]{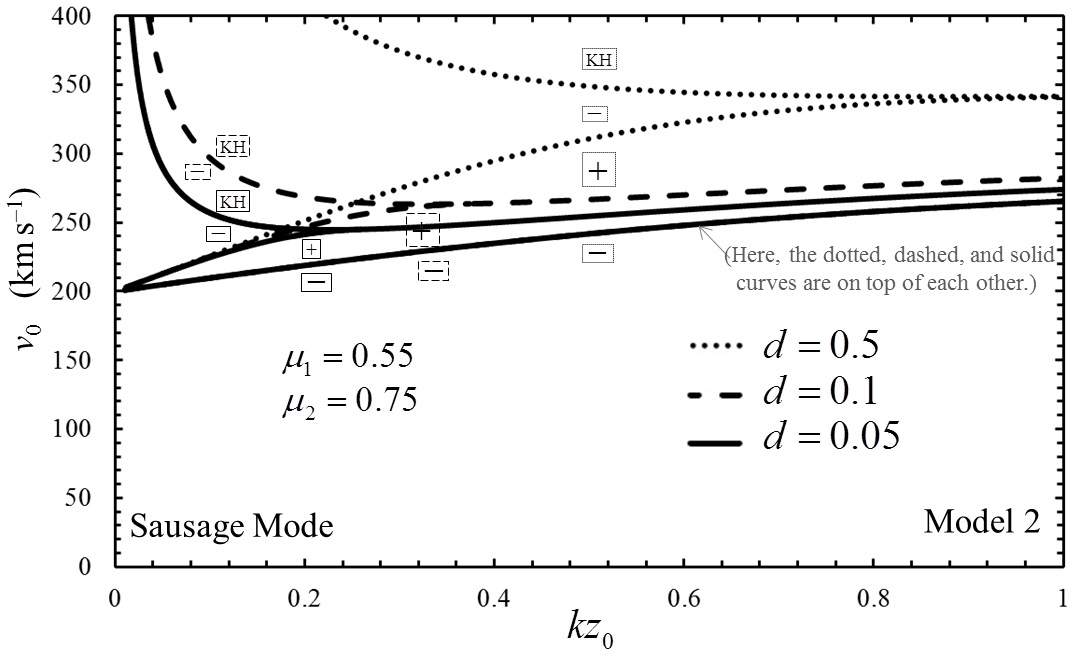}
\caption{Contour plot of the variation of $\omega_{i2}$ in the case of sausage modes in terms of the equilibrium flow speed and the value of the dimensionless parameter $kz_0$ for model 2. The values of the density is shown in the legend of the figure. The region where instability occurs is shown by the {\it plus} symbol.}
\label{fig6}
\end{figure}
\begin{figure}
\centering
\includegraphics[width=\columnwidth]{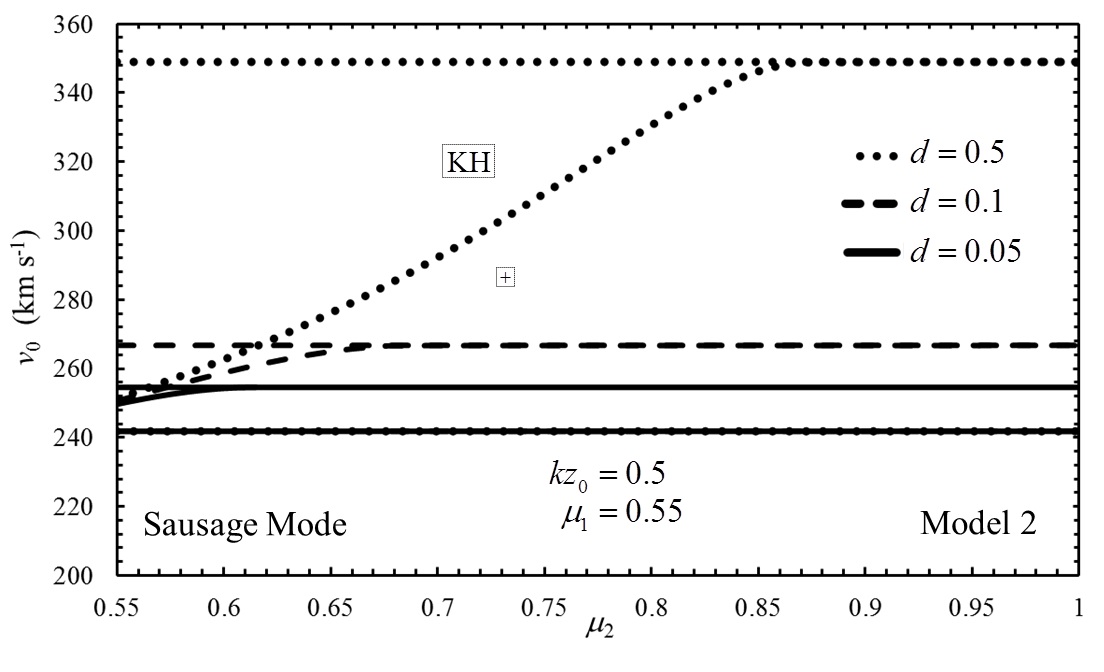}
\caption{Contour plot of the variation of $\omega_{i2}$ in the case of sausage modes in terms of the equilibrium flow speed and the ionisation degree of the external region for model 2. The values of the dimensionless quantity $kz_0$ and the ionisation degree of the plasma slab has been fixed and shown on the figure. The region where instability occurs is shown by the {\it plus} symbol.}
\label{fig6.1}
\end{figure}
First to note is that the values of the equilibrium flow at which the backward mode is unstable for all three values of the density contrast are far too high. For flow spends that are observed modes are stable and they encounter a normal physical damping. Backward propagating waves become unstable for flow speeds that are comparable with the KH speeds. In Fig. 5, backward propagating waves are unstable only in a narrow band shown by a {\it plus} sign. Above the upper boundary of the instability zone, the plasma becomes KH unstable. Similar findings can be obtained for other density contrasts. 
\begin{figure}
\centering
\includegraphics[width=\columnwidth]{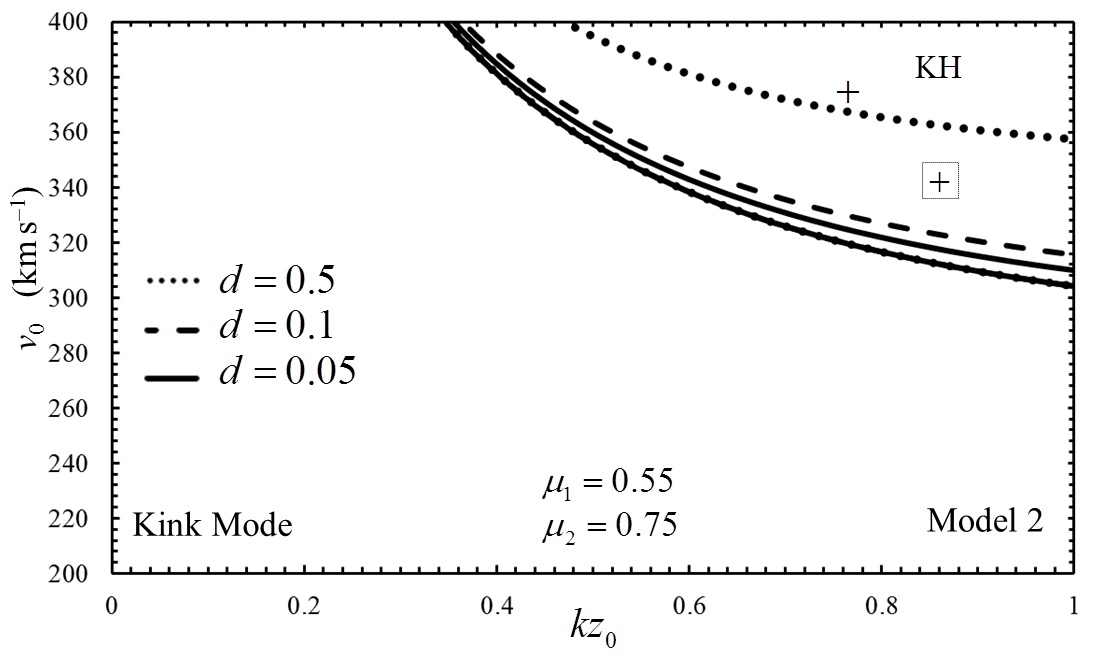}
\caption{Similar to Fig. 5, but here we plot the regions of instability for kink waves.}
\label{fig7}
\end{figure}
In Fig. 6, the appearance of the unstable regime of the same sausage mode is displayed for a fixed value of $kz_0$ and we allow the variation of the ionisation of the external medium to change between the extreme cases of a near complete ionisation and near neutral gas. Similar to the findings of Fig. 5, the values of flows at which the instability sets in is at around 250 km s$^{-1}$, clearly for values higher than any measured flow speeds. Similar to the results found in Fig. 5, the instability appears only in a limited region, that is a well-specified combination of parameters. For $d=0.05$, the plasma is unstable in a very small region for values of the external ionisation, that is, close to full ionisation. Everywhere else, waves propagating in the slab will damp. 

In the case of kink waves (see Fig. 7), the instability sets in, again, for very large values of the equilibrium flows and for very long wavelengths. The unstable region, again, is bounded, similarly to the domain shown in Fig. 4.
Finally, for the same kind of waves, the instability threshold is practically independent on the ionisation degree (for a fixed value of $kz_0$).

\noindent The very large values of equilibrium flows at which dissipative instabilities occur in filamentary structures lead us to the necessity of an alternative treatment of the partially ionised prominence plasma stability. 

\subsection*{Two-fluid approximation}

The above one fluid models are restricted to periods that are larger than the ion-neutral collisional time. For periods that are less or comparable to the ion-neutral collisional time, the plasma dynamics is described in a  two-fluid approximation, where the equations are written for the charged particles (ions and elections) and neutrals. The plasma behaviour of the mixture is ensured by collisional terms between ions and neutrals in the induction equation and these collisional are dominant processes in momentum transfer between species and the electron inertia is neglected. In reality, the two models are not identical, instead they are complementary.

Similarly to the equilibrium described earlier in this study (model 2), we are going to consider that the equilibrium configuration is composed of a partially ionised slab filled with plasma in steady state (${\bf v}_0$ is the equilibrium flow and is parallel to the discontinuity), surrounded by two partially ionised half-space plasma regions. The plasma is permeated by a homogeneous magnetic field oriented in the $x$ direction, and the interfaces are situated at $z=0$ and $z=z_0$. The equilibrium plasma parameters are homogeneous and constants in all regions. We denote the regions inside and outside the slab by the indices 1 and 2, respectively. 

The set of coupled differential equations governing the dynamics of linear waves in incompressible two-fluid plasmas is given by (see, e.g. Zaqarashvili et al. 2011, Khomenko et al. 2014a, Mart\'inez-G\'omez et al. 2015 )
\[
\rho_{0i}\left(\frac{\partial {\bf v}_i}{\partial t}+{\bf v}_0\cdot \nabla {\bf v}_i\right)=-\nabla p_{ie}+\frac{1}{\mu_0}(\nabla\times {\bf b})\times {\bf B}_0-
\]
\begin{equation}
-\alpha_{in}({\bf v}_i-{\bf v}_n),
\label{eq:4.51}
\end{equation}
\begin{equation}
\rho_{0n}\left(\frac{\partial {\bf v}_n}{\partial t}+{\bf v}_0\cdot \nabla {\bf v}_n\right)=-\nabla p_{n}-\alpha_{in}({\bf v}_n-{\bf v}_i),
\label{eq:4.52}
\end{equation}
\begin{equation}
\frac{\partial {\bf b}}{\partial t}=\nabla\times ({\bf v}_0\times {\bf b})+\nabla\times ({\bf v}_i\times {\bf B}_0)+{{\bf \cal R}_1},
\label{eq:4.53}
\end{equation}
\begin{equation}
\nabla\cdot {\bf v}_i=\nabla\cdot {\bf v}_n=\nabla\cdot {\bf b}=0,
\label{eq:4.54}
\end{equation}
where ${\bf v}_i = (v_{ix}, 0, v_{iz})$ and ${\bf v}_n = (v_{nx}, 0, v_{nz})$ are the components of the two-dimensional velocity
perturbation of ions and neutrals, $p_{ie}$ and $p_n$ are the pressure perturbations of the ion-electron and neutral fluids, ${\bf b} = (b_x, 0, b_z)$ is the magnetic field perturbation, $\rho_{0i}$ and $\rho_{0n}$ are the equilibrium densities of ions and neutrals, while $\alpha_{in}$ is the ion-neutral friction coefficient. Frictions between charged and neutral (close-range interaction) particles is ensured via collision processes. In the absence of this process, neutrals will not be able to stay in the system and the momentum equations would decouple. Equations (\ref{eq:4.51}) and (\ref{eq:4.52}) are the linearized momentum equations of the ion-electron fluid and neutrals, respectively. The last terms on their RHSs express the transfer of momentum between ions and neutrals through diffusion of one species into the other.  As a result of collisions, particles can loose energy. The dynamics in the external region is described by a similar system of equations, with the exception that the equilibrium outside the slab is static. As pointed out by Zaqarashvili et al. (2011), Cowling resistivity appears only in the one fluid approximation, that is why in Eq. (\ref{eq:4.53}) ${\bf \cal R}_1=\eta\nabla^2{\bf b}$, where the coefficient of resistivity appears only due to the movement and interaction of electrons in the partially ionised plasma and is defined as
\[
\eta=\frac{c^2(\nu_{ei}+\nu_{en})}{\omega_{pe}^2},
\]
where $\nu_{ei}$ and $\nu_{en}$ are the electron-ion and electron-neutrals collisional frequencies and $\omega_{pe}$ is the electron plasma frequency. According to Braginskii (1965) and Zaqarashvili et al. (2011) the two collisional frequencies are defined as
\[
\nu_{ei}=\frac{\sqrt{12}n_ee^4\ln \Lambda}{12\pi^{3/2}\epsilon_0^2m_e^{1/2}(k_BT)^{3/2}}, \;\;\; \nu_{en}=\Sigma_{en}n_n\left(\frac{8k_BT}{\pi m_n}\right)^{1/2},
\]
where $n_e$ is the number density of electrons, $e$ is the electron charge, $\ln \Lambda$ is the plasma logarithm, $\epsilon_0$ is the permitivity of free space and $\Sigma_{en}=10^{-19}$ m$^2$ is the electron-neutral collision cross-section. The electron plasma frequency is a quantity that depends only on the number density of electrons and is defined as $\omega_{pe}\approx 17.8\pi n_e^{1/2}$ (s$^{-1}$). Introducing the values of physical constant and assuming $\ln \Lambda \approx 15$ we arrive to the relation
\[
\eta=\left[\frac{4.95(1-\mu)T^{1/2}}{2\mu-1}+\frac{15.81\times 10^8}{T^{3/2}}\right] \quad (m^2 s^{-1}),
\]
where the temperature is measured in $K$. Equation (\ref{eq:4.53}) clearly shows that the magnetic field is able to interact only with the charged part of the plasma fluid. 

Let us express the ion-neutral friction coefficient, $\alpha_{in}$, as 
\begin{equation}
\alpha_{in}=\rho_{0i}\rho_{0n}\gamma_{in},
\label{eq:4.56}
\end{equation}
where $\gamma_{in}$ is the ion-neutral collision rate coefficient per unit mass. The friction coefficient vanishes in both the fully ionized ($\rho_{0n}=0$) and fully neutral ($\rho_{0i}=0$) cases. However, instead of using $\gamma_{in}$, we are going to use the collision frequency, which has a more practical physical meaning. Thus, we define the ion-neutral, $\nu_{in}$, and neutral-ion, $\nu_{ni}$, collision frequencies as
\begin{equation}
\nu_{in}=\rho_{0i}\gamma_{in}, \qquad \nu_{ni}=\rho_{0n}\gamma_{in},
\label{eq:4.57}
\end{equation}
and the two collisional frequencies are connected through $\rho_{0n}\nu_{in}=\rho_{0i}\nu_{ni}$. In consequence, in the remaining part of the present paper, we are going to use $\nu_{in}$ only and use $\nu_1$ and $\nu_2$ to denote the collisional frequencies in the two media. 

We are going to employ the same normal mode analysis as in the case of single fluid description here, however, the continuity of the normal component of momentum across the discontinuity would require an equivalent relation written for ions and neutrals. Similarly, the continuity of the stresses at the interface would translate into the balance of the total pressure of charged particles and the kinetic pressure of neutrals.

After straightforward calculations we can obtain that the dispersion relation for waves propagating along the interface in the incompressible limit can be given as ${\cal D}_R+i{\cal D}_I=0$, where now the real and imaginary parts of the dispersion relation are given by
\[
{\cal D}_R=(D_{A1}d_i+D_{A2}\tanh kz_0)(\Omega^2 d_n+\omega^2 \tanh kz_0),
\]
\[
{\cal D}_I=(D_{A1}d_i+D_{A2}\tanh kz_0)(\Omega\nu_{1}d_n+\omega\nu_{2}\tanh kz_0)+
\]
\[
(\Omega^2 d_n+\omega^2 \tanh kz_0)\left[\Omega d_i\nu_{1}\chi_1+\omega\nu_{2}\chi_2\tanh kz_0+\right.
\]
\begin{equation}
\left.+k^4\left(\frac{v_{A1}^2\eta_{1}d_i}{\Omega}+\frac{v_{A2}^2\eta_{2}\tanh kz_0}{\omega}\right)\right].
\label{eq:4.59}
\end{equation}
where $D_{A1}=\Omega^2-k^2v_{A1}^2$, $D_{A2}=\omega^2-k^2v_{A2}^2$, $\chi_{1,2}=\rho_{0n1,2}/\rho_{0i1,2}$ and $d_n=\rho_{0n1}/\rho_{0n2}$, $d_i=\rho_{0i1}/\rho_{0i2}$. The dispersion relation given by Eq. (\ref{eq:4.59}) describes the propagation of two pairs of waves propagating in the opposite direction, in each direction having a wave that is connected to ions, while the other one to neutrals. The collisions between species lead to the modification in the amplitude of waves. The presence of collisions between particles (and the associated momentum transfer) together with the resistivity renders the equations to be dissipative. Due to the loss of energy and momentum of individual species waves will propagate with a complex frequency, where the imaginary part describes damping or amplification. 

Despite lacking a firm physical basis from the partially ionised plasma point view, let us discuss the collisionless and ideal limit, that is when $\nu_{1}=\nu_{2}=\eta_{1}=\eta_{2}=0$ (in the absence of collisions, neutrals cannot be kept in the system), as this will help us understand the results obtained in the collisional limit. In this case, the dispersion relation is decoupled and we can solve separate equations for ions and neutrals. Accordingly, the dispersion relation for ions becomes
\begin{equation}
D_{A1}d_i+D_{A2}\tanh kz_0=0,
\label{eq:4.58.1}
\end{equation}
which can be easily solved to lead to 
\begin{equation}
\omega=k\frac{v_0d_i\pm\sqrt{d_i\tanh kz_0(v_{KH}^2-v_0^2)}}{d_i+\tanh kz_0},
\label{eq:4.58.2}
\end{equation}
where the Kelvin-Helmholtz speed is defined as 
\[
v_{KH}^2= \frac{(d_i+\tanh kz_0)(v_{A1}^2d_i+v_{A2}^2\tanh kz_0)}{d_i\tanh kz_0}.
\]
As Eq. (\ref{eq:4.58.2}) shows, the ion wave becomes KH unstable for flow speeds larger than $v_{KH}$. However, since $v_{KH}$ is always super-Alfv\'enic in an incompressible plasma and the observed flows are always sub-Alfv\'enic, waves due to ions will be always KH stable. 

In the case of neutrals, the dispersion relation in the collisionless limit becomes
\begin{equation}
\Omega^2 d_n+\omega^2 \tanh kz_0=0,
\label{eq:4.58.3}
\end{equation}
whose solution reads
\begin{equation}
\omega=k\frac{v_0d_n\pm iv_0\sqrt{d_n\tanh kz_0}}{d_n+\tanh kz_0}.
\label{eq:4.58.4}
\end{equation}
This dispersion relation describes the propagation of two waves in the same direction, however, one of them is damped, while the other is amplified in time. These modes owe their existence to the presence of the equilibrium flow, and the flow plays the role of reservoir/sink for gained/lost energy of the wave. The peculiar behaviour of neutrals under prominence conditions was discussed earlier (see, e.g. Soler et al. 2012) and the unstable behaviour corresponds to the standard hydrodynamic KH instability. Another important result of this limit is that a two-fluid approach allows the propagation of two pairs of waves, one due to charged particles, and the other one for neutrals. In contrast, a single-fluid approach allows the propagation of one single pair, and this mode is a surface Alfv\'en wave. Similarly to the single-fluid approach, we are interested in the backward propagating waves that can develop dissipative instability. 

Let us first concentrate on the ion-related waves, for which the dispersion relation (in the uncoupled limit) of the backward propagating wave is given by Eq. (\ref{eq:4.58.2}), where we choose the lower sign.
Using Cairn's criteria, the imaginary part of the frequency is approximately given by Eq. (\ref{eq:3.1}) where the equation is evaluated at the roots of the ideal dispersion relation, that is the solutions obtained in the collisionless limit.  

After straightforward calculations we obtain that the imaginary part of the frequency of the backward propagating wave is given by 
\begin{equation}
\omega_{ii}\approx \frac{\theta_1+\theta_2}{2 \tilde{\Gamma}_1(d_i+\tanh kz_0)},
\label{eq:4.61}
\end{equation}
where 
\[
\theta_1=\frac{k^2(d_i+\tanh kz_0)^2}{(v_0d_i-\tilde{\Gamma}_1)(\tilde{\Gamma}_1+v_0\tanh kz_0)}\left[\tilde{\Gamma}_1(v_{A1}^2\eta_{1}d_i+v_{A2}^2\eta_{2})+\right.
\]
\[
+\left.v_0(v_{A2}^2\eta_{2}\tanh kz_0-v_{A1}^2\eta_{1}d_i^2)\right]
\]
\[
\theta_2=- \tilde{\Gamma}_1(d_i\nu_{1}\chi_1+\nu_{2}\chi_2\tanh kz_0)-
\]
\[
v_0\tanh kz_0d_i(\nu_{1}\chi_1-\nu_{2}\chi_2),
\]
and $ \tilde{\Gamma}_1=\sqrt{d_i\tanh kz_0(v_{KH}^2-v_0^2)}$.

Let us apply this relation in connection to the modes that could appear in prominence dark plumes. A higher temperature in the slab would mean that more neutrals are ionised, therefore, $d_n<1$. The ion-neutral collisional frequency depends on the temperature and densities of the plasma. Let us consider that $\nu_{1}=\alpha\nu_{2}$. As a result, the ratio of the collisional frequencies becomes
\[
\alpha=\frac{\nu_{1}}{\nu_{2}}=\frac{\rho_{0n1}}{\rho_{0n2}}\sqrt{\frac{T_1}{T_2}}=d_n\sqrt{\frac{T_1}{T_2}}.
\]
Although we restricted ourselves to the case $d_n<1$, the fact that $T_1/T_2>1$ means that $\alpha$ can take any values. As a representative collisional frequency between ions and neutrals in the prominence region we can estimate that the ion-neutral collisional time (for a hydrogen plasma) can be defined as
\begin{equation}
\nu_{in}=4n_n\Sigma_{in}\sqrt{\frac{k_B T}{\pi m_i}}.
\label{eq:4.64.1}
\end{equation}
For typical prominence values ($n_n= 10^{16}$ m$^{-3}$, $T=10^4$ K using the FAL3 model), we obtain a collisional frequency of the order of 208 s$^{-1}$ (Fontenla et al. 1990, Oliver et al. 2016). 

To bring theoretical results closer to observations, we express neutral densities in terms of ion densities, as current observations reveal number densities for ions, rather than neutrals. From Eqs. (\ref{eq:3.8})-(\ref{eq:3.10}) we have
\[
\rho_{0n1}\approx \frac{\xi_{n1}\rho_{0i1}}{1-\xi_{n1}},
\]
and a similar relation is valid in region 2. It is easy to calculate the new expression of $d_n$ as
\[
d_n=d_i\frac{(2\mu_1-1)(1-\mu_2)}{(1-\mu_1)(2\mu_2-1)},
\]
where $\mu_1$ and $\mu_2$ are the ionisation fractions inside and outside the slab, respectively.

Let us now consider that medium 1 is a plume and region 2 represents the surrounding prominence. Based on observations by Berger et al. (2010), we assume that plumes, having a temperature of $10^5$ K, are surrounded by prominence plasma with temperature of $10^4$ K. Since the exact ionisation degree of the prominence plasma is not known, we assume for illustration that $\mu_2=0.75$ and let the ionisation degree of the plume vary between 0.5 and 0.75. Furthermore, we consider that the ion density in the prominence is $\rho_{i02}=5\times 10^{-11}$ kg m$^{-3}$ and a density of the plume is 10\% of it, that is $\rho_{0i1}=0.1\rho_{0i2}$, $d_i=0.1$. For all types of waves, we adopted a flow speed of 10 km s$^{-1}$. 
\begin{figure}
\centering
\includegraphics[width=\columnwidth]{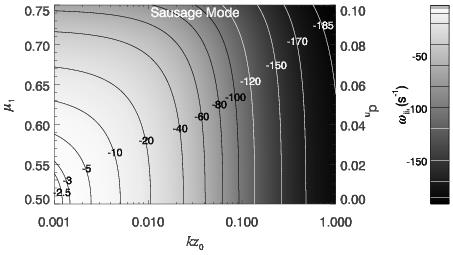}
\caption{Variation of the imaginary part of the frequency for backward propagating sausage modes due to ions in the two-fluid approximation with the ionisation degree of the plasma and the wavelength of waves.}
\label{fig9}
\end{figure}
\begin{figure}
\centering
\includegraphics[width=\columnwidth]{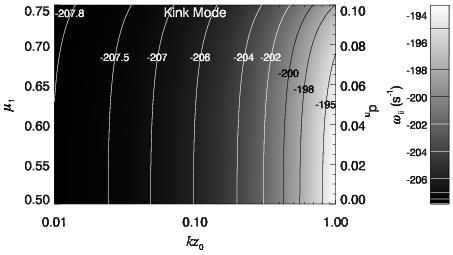}
\caption{The same as in Fig. 8, but here the dispersion curves denote the imaginary part of the frequency of kink modes.}
\label{fig10}
\end{figure}
Since the imaginary part of the frequency is negative for both waves (regardless the ionisation degree), the waves will undergo a physical damping (Figs. 8--9). Similar to the one fluid model, the instability will set in for much larger values of flow, values that are currently not observed, however the more ionised the plasma is, the easier is to have an unstable sausage wave. From this perspective, kink modes are more stable than sausage modes.

Now let us look at the waves associated to neutrals where both waves are forward propagating. Now, if unstable behaviour appears, it could be attributed to a Kelvin-Helmholtz instability (similar to the findings by Mart\'inez-G\'omez et al. 2015). Since the mode corresponding to the lower sign is always damped, we will concentrate on the wave that corresponds to the upper sign in Eq. (\ref{eq:4.58.4}). In the collisionless limit the amplitude of this wave grows in time. We should mention here that the imaginary part of the frequency given by Eq. (\ref{eq:4.58.4}) does not refer to any physical effect. This value is the solution of the collisionless limit and it will be used next to determine the real (physical) damping or growth rate of these waves. Following the same procedure as before, the imaginary part of the frequency is given by Eq. (\ref{eq:A1}) and it is clear that it is not affected by resistivity. 

However, the expression given by Eq. (\ref{eq:A1}) is not the full answer to our problem, since this has to be considered together with the value determined earlier in Eq. (\ref{eq:4.58.4}). Combining the two relations, we arrive at the imaginary part of the frequency that appears due to neutrals
\begin{equation}
{\omega}_{in}^{+}=\overline{\omega}_{in}^{+}+\frac{kv_0\sqrt{d_n\tanh kz_0}}{(d_n+\tanh kz_0)}.
\label{eq:4.64}
\end{equation}
The variation of the imaginary part of the frequency (counterplots) is shown in Figs. (10)--(11), for sausage and kink modes, respectively.
\begin{figure}
\centering
\includegraphics[width=\columnwidth]{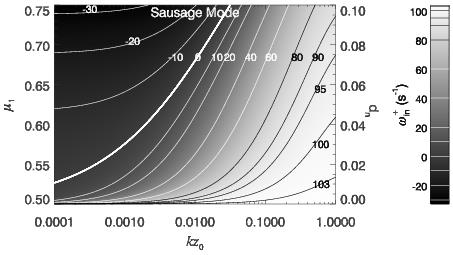}
\caption{Variation of the imaginary part of the frequency for propagating sausage modes in the two-fluid approximation.}
\label{fig11}
\end{figure}
\begin{figure}
\centering
\includegraphics[width=\columnwidth]{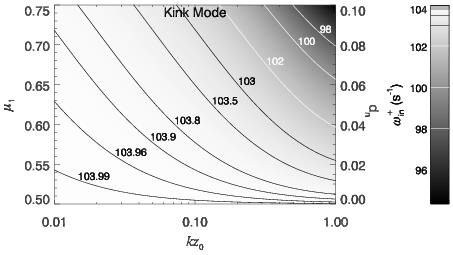}
\caption{The same as in Fig. 10, but here the dispersion curves denote the imaginary part of the frequency of kink modes.}
\label{fig12}
\end{figure}
This result clearly shows what new physics is brought into these systems by neutrals. While kink waves are unstable for the whole range of parameters, the instability threshold for sausage modes depends on the ionisation degree of the plasma and the wavelength of waves (Fig. 10). In the long wavelength limit ($kz_0\ll 1$) waves are unstable until the concentration of neutrals in the plume regions reaches a certain threshold value, after which these modes will damp. The value of the critical neutral concentration increases with decreasing the wavelength of waves. In addition, there will be a wavelength cut-off value (here corresponding to $kz_0\approx 0.03$) after which modes will become unstable, regardless the amount of neutrals in the system. In the case of sausage modes the growth rate of waves (for fixed wavelength) decreases with increasing the number of neutrals in the plume region. The same pattern is recovered for kink modes, however, the growth rate is weakly affected by the presence of more neutrals. Therefore, neutrals indeed have a stabilising effect on sausage modes due to the increased amount of the momentum transferred by neutrals in the process of collision. For a fixed ionisation degree, the frequency increases with decreasing wavelength, so that the  longer the wavelength, the shorter is the growth time of sausage waves.  

These results show that the consideration of a two fluid MHD for model 2 introduces a new physics that has not been possible to describe in the framework of a single fluid MHD. The two regimes are difficult to compare, as they are valid in different frequency regimes. It remains to be seen what the temporal evolution of these instabilities is (e.g. they can saturate, evolve into a macro instability, or develop turbulences), but this issue would require a robust numerical investigation. The mixture of two species plasma, therefore, shows a very complex pattern where the Kelvin-Helmholtz unstable behaviour of neutrals is stabilised by the damping due to collisions with ions. Since the species of the plasma are coupled through collisions, the instability of one single species will drive the whole mixture into an unstable state. 

Finally, we need to mention that our results are somehow quantitatively different from the results obtained earlier by Mart\'inez-G\'omez et al. (2015), but this is due to the technique used here to find the imaginary part of the frequency. While the above authors used numerical methods to study the onset of the KHI in a two-fluid model, our analytical approach is valid only for weakly dissipative and low collisional frequency plasmas. 

\section{Conclusions}

The present research focussed on the appearance of dissipative instabilities for waves propagating in a partially ionised plasma slab surrounded by the corona or another partially ionised prominence environment. The geometrical restrictions imposed on waves make them dispersive, and different characteristics were investigated for symmetric and asymmetric waves (sausage and kink waves). 

The nature of the instabilities discussed here means that they appear for flow speeds lower than the KHI, the value of the KH speed playing a special role in our discussion. A simple analysis showed that the KHI is unlikely to occur in the plasmas we dealt with. The threshold values where the KH instability occurs varies with the density ratio of the slab plasma and its surrounding and with the wavelength of the waves. In all cases, the flow speed at which waves are KH unstable is of the order of a few thousand km s$^{-1}$, that is much above currently observed values.

We analysed the appearance of dissipative instability, that is the unstable growth of a backward propagating wave in the presence of flow, in two different equilibrium set-ups. First we assumed that the whole partially ionised prominence can be treated as a slab surrounded by the viscous and completely ionised corona. After imposing the necessary boundary conditions at the interfaces between the two media, we derived a dispersion relation describing the propagation of incompressible waves propagating inside the slab. The imaginary part of the frequency describes the damping or the growth of waves. The results on the role of viscosity and magnetic field are identical with the findings by Ballai et al. (2015). Here we focussed on the role of dispersion and the ionisation degree on the stability threshold of waves. Our results show that sausage modes are more sensitive to the variation of physical parameters, the value of ionisation degree is more pronounced for very large wavelengths. 

The second model we used is a plasma slab in partially ionised state surrounded by another partially ionised, infinitely extended environment, modelling the case of prominence dark plumes surrounded by another prominence material.  Our results clearly show that the unstable behaviour of these structures require unrealistically high flows as long as the periods of waves are larger than the ion-collisional time and the plasma dynamics is described in a one-fluid MHD model. The situation is different when we consider very high frequency waves (or low periods) when the plasma behaviour requires a two-fluid approach. 

This investigation also shows the complexity of the physical situation under investigation and the importance of a two-fluid MHD approach in partially ionised plasmas. In this description, the plasma becomes unstable solely due to the propagating waves that appear in the presence of neutrals. The source of instability for this mode in a single-fluid MHD approach does not exist. Strictly speaking the instability described here is not dissipative in the sense of instabilities described earlier, as the modes that are unstable are forward propagating.

Large values of flow necessary for an instability to occur in a single fluid MHD model is not a surprise, given the simplicity of our model. We are aware that the simple model employed here misses several key effects for plume dynamics. For example, the sharp interface considered here does not allow the appearance of body waves, that could be important concerning the stability of plumes. Furthermore, the flow of particles oblique to the magnetic field might decrease the instability threshold significantly, as it was shown by Prialnik et al. (1986) in the case of KH instability. It also remains to be seen how the effect of compressibility can change the stability criteria, bearing in mind that the general effect of compressibility is to stabilise the plasma.

\acknowledgements

The authors are grateful for the support by the Leverhulme Trust (IN-2014-016). R.O. acknowledges support from MINECO and FEDER funds through project AYA2014-54485-P. The authors would like to thank R. Soler, D. Mart\'inez-G\'omez, and the anonymous referee for their suggestions

\appendix

\section{The imaginary part of the frequency calculated for neutrals}

The imaginary part of the frequency calculated with the help of Cairn's formula (see Eq. (\ref{eq:3.1}) that corresponds to the positive (amplified) solution of the collisionless dispersion relation attached to neutrals is given by
\begin{equation}
\overline{\omega}_{in}^{+}=-\frac{1}{2(d_n+\tanh kz_0)}\left(\nu_{1}d_n-\nu_{2}\tanh kz_0+\frac{\Psi_1^{+}}{\Psi_2^{+}}\right),
\label{eq:A1}
\end{equation}
where
\[
\Psi_1^{+}=2d_n\tanh^2 kz_0(\nu_{1}-\nu_{2})\frac{v_0^2(d_i-d_n)}{(d_n+\tanh kz_0)^2},
\]
\[
\Psi_2^{+}=\frac{v_0^2(d_i-d_n)\tanh kz_0}{(d_n+\tanh kz_0)^2}(\tanh kz_0-d_n)-v_{A1}^2d_i-v_{A2}^2\tanh kz_0.
\]

\end{document}